\documentclass[12]{article}
\usepackage{amssymb}

\usepackage{graphics} 
\usepackage{epsfig}

\usepackage{amsmath}

\begin{document}
\begin{center}
{\bf \Large A projection operator technique for solution of
relativistic wave equation on  non-compact group: the case of  a
charged vector-boson
 }\\[0.2cm]
 {Halina V. Grushevskaya$^{1}$ and Leonid I. Gurskii$^{2}$
}\\[0.2cm]
 {\it $^{1}$ Physics Department, Belarusian State University,\\
 4 Fr. Skorina Av., 220050 Minsk, Belarus \\
 $^{2}$ Belarusian State
 University of Informatics  and Radio electronics,\\
 6 P. Brovka Str, 220027 Minsk, Belarus}\\[0.2cm]
{E-mail: grushevskaja@bsu.by}
\end{center}

\begin{abstract}
A projection operator technique for solution of relativistic wave equation on
  non-compact group has been  proposed.  This technique was applied to the
construction of wave equations for  charged vector boson in a
potential field. The
equations were shown to approximately describe a hydrogen-like atom and allow
estimating of relativistic corrections such as a fine structure of
hydrogen atom lines with  high accuracy.
\\[0.2cm]
\end{abstract}

\section{Introduction}

Geometrical approach plays an important  role in the study of
quantum models and   relativistic processes
\cite{Volobiuev,{Gurskii1},{Gurskii2}}.
 This paper discusses the projection operator technique to examine motion of a
relativistic vector boson with a non-compact
symmetry group \textsf{SO(4,2)}. The constructed equations of
motion can also be used for the description of a
hydrogen-like atom, since being a composite quantum system, it has an
integer-valued spin. In the paper, these
equations are shown to estimate relativistic corrections such as a fine structure of
hydrogen atom lines with  high accuracy.

The goal  of the paper is to develop a geometrical method of the
projection operators and on this base to examine motion of charged vector
boson in a potential field.

\section{Projection operator technique}
Let us consider an equation for a relativistic quantum particle in
two-dimensional flat space with two complex coordinates
\begin{eqnarray}
|\xi_s \rangle , |\xi_{\dot{s}
} \rangle ,\qquad
 s(\dot{s}
 )=1,2 ; \label{Weyl-realization}
\end{eqnarray}
being components of ket bispinor.

Consider motion of a free particle, then the
Hamiltonian $ \cal H $ includes only relativistic kinetic energy
term  $T $. For comparison, we shall remind a classic analog $T^2=Z^2
+ {\vec k}
^2 $. 
Here $ \vec k $ is a momentum, $Z $ is a "mass", $c $ is the
velocity of light:  $c=1 $. Then, a relativistic wave equation
reads
\begin{eqnarray}
T|\psi\rangle = \imath {\partial |\psi \rangle \over \partial t}
\label{fist-order-rel-eq}
\end{eqnarray}
where $|\psi\rangle$ is a wave function,
$t$ is  time, $\imath $ is imaginary unit, $\hbar =1$.
Eq. (\ref{fist-order-rel-eq}) can be rewritten as
\begin{eqnarray}
\left({\cal H}^{\dag}{\cal H}+ {\partial^2 \over \partial t^2}
\right) |\psi\rangle  =\left( -{\partial ^2\over \left.\langle
\partial\xi 
_{\dot{s}} \right| \partial \xi _s\rangle } -Z^2 + {\partial
^2\over \partial t^2} \left|\xi _{\dot{s}}\rangle \langle \xi _s\right|
\right)\psi\rangle \equiv ( H-Z^2 )|\psi\rangle = 0
\label{rel-eq}.
\end{eqnarray}
Here the summation is assumed over repeated indexes, and it has
been taken into account that the following expression holds:
$|\xi_{\dot s}
\rangle \langle \xi_s | = \hat I $,
where $\hat I$ is the operator unity;
the wave function
$ \left < \xi_s | \psi \right > $ is considered in the representation
$ | \xi_s \rangle $, $s=1,2 $.
It follows from Eq.~(\ref {rel-eq}) that it is possible to
consider the square of   $Z $ as eigenvalues of the operator $H $,
which is equal to $ \imath {\partial \over \partial t_1} $, where
$t_1 $ is a time coordinate which does not depend  on $t $. Hence,
the equation (\ref {rel-eq}) is determined on the manifold with four
space-like and two time-like  coordinates.

Let us pass  into the representation $ \left \{\left.\left |
\phi_i\right. \right > \right \} $ where the operator
representation $ \hat \xi\equiv T_\xi $ of a translation group   $
\{\hat \xi \} $ has the following form: $ \hat \xi \to \imath {\partial
\over
\partial k_i} $,  $ k_i $ is a momentum. Then, using
the properties of the projection operators \cite{barut}
Eq.~(\ref{rel-eq}) can be rewritten as
\begin{eqnarray}
 \left(
-{\partial ^2\over \left.\langle \partial\xi 
_{\dot{s}} \right| \partial
\xi _s\rangle } -Z^2 + {\partial ^2\over \partial t^2} \delta_{ki}
\left( \left|\phi \rangle \langle \phi _k\right| \xi _{\dot s}\rangle
\langle \xi _s \left|\phi _i\rangle \langle \phi _r\right| \right)
\right)\psi_r\rangle  = 0. \label{rel-eq1}
\end{eqnarray}
The components of  the wave function for the free particle in this
representation are
\begin{eqnarray}
\left<\phi_i|\psi\right> = \exp\left\{\imath \left( k_s\xi_{si}
+\epsilon_{srq}k_s {1\over |\xi|^2 }[\xi_{ri},\xi_{qi}] \right) \right\}
\exp(\imath \omega t),\\
\xi_{si}\equiv <\phi_i|\xi_s> . \label{presentation}
\end{eqnarray}
Here the first term of the expression (\ref{presentation})
describes the translation  to the group unity; the commutator appeared
in the second summand is an element of the rotation group;
 $\omega $ is a frequency, $ | \xi | ^2 =\sum _ {\nu =1} ^4
\xi^2_\nu $, $ \xi _ {si} $ are components of bispinor in
representation $ \{\phi_i\}_{i=1} ^N $, $N=1,2, \ldots $, being
$c  $-numbers; $ \epsilon _ {ijk} $ is the Levi-Civita tensor.

Using the definition of a projector, after differentiation with respect to
$t, \vec k $, we transform the equation (\ref {rel-eq1}) to the
following form
\begin{eqnarray}
 \left(
-{\partial ^2\over \left.\langle \partial\xi 
_{\dot{s}} \right| \partial
\xi _s\rangle } -Z^2 + \omega ^2 \left( \xi_{{\dot s}i}
+\epsilon_{\dot s \dot r \dot q}
[\xi_{\dot r i},\xi_{\dot q i}] \right) \left( \xi_{si}
+\epsilon_{skq}[\xi_{ki},\xi_{qi}] \right) \left|\phi _i\rangle
\langle \phi _r\right|
 \right)\psi _r\rangle  = 0
\label{rel-eq2}
\end{eqnarray}
Assuming that  the basis $ \{\xi _ {si} \} $ obeys a condition of
orthogonality,   finally  Eq.~(\ref {rel-eq2}) is written in the
reduced form
\begin{eqnarray}
 \left(
-{\partial ^2\over \partial \vec \xi 
_{\dot{s}}  \partial \vec \xi _s }
-Z^2 + \omega ^2 \left( \vec \xi_{\dot s} \vec \xi_{s} +
\epsilon_{\dot s \dot r \dot q}{1\over |\xi|^4 }
[\vec \xi_{\dot r},\vec \xi_{\dot q}]
\epsilon_{srq}[\vec \xi_{r},\vec \xi_{q}] \right) \left|\Phi
\rangle \langle \Phi \right|
 \right)\psi\rangle  = 0
\label{rel-eq3}
\end{eqnarray}
where $\vec \xi_s\stackrel{ def}{=\hspace{-0.8mm}=}\{\xi_{si}\}_i$
 are the components of bispinor in the basic set
$\Phi\stackrel{def}{=\hspace{-0.8mm}=}\left\{\left.\left|
\phi_i\right. \right>\right\}$.

To  find a real  equation of motion, let us
 consider particle motion in  three-dimensional space. It
is known \cite{Giemin} that the mapping of  three-dimensional
sphere $S^3\in \mathbb {R} ^4 $ into  three-dimensional space $
\mathbb {R} ^3 $ is two-valued. Therefore, the set of four numbers
describing a position of a particle in  four-dimensional space,
is replaced by a set  of four numbers $ \{x_\lambda, S \},\
\lambda=1,2,3 $ also; but one of them ($S $) is a discrete quantity. $S $ takes two values $S =\pm 1 $ depending on either
we exclude north  or south pole when projecting. The physical
interpretation of $S $ is the helicity  of  a particle
as a result of the projection $P $ on three-dimensional
subspace. For the ket vector $ |x_\lambda ^ {R (L)} \rangle $ the
symbols $R, L $ are attributed to right- and left-helical particles,
respectively. Then, we can write the following identity
\begin{eqnarray}
\sum_\mu |\xi_\mu \left> \right< \xi_\mu|\psi \rangle + {1\over
2}\sum_\lambda \left[ |\xi_\kappa \left> \left(\sigma^+_\lambda
\right)_{\kappa \nu} \right< \xi_\nu|\psi \rangle+ |\xi_\kappa
\left> \left(\sigma^-_{\lambda} \right)_{\kappa \nu} \right<
\xi_\nu|\psi \rangle \right]
\nonumber \\
\equiv \sum_{\lambda}\left[ {1\over 2}\left( |x_\lambda ^{L}
\rangle \langle x_\lambda ^{L} |+ |x_\lambda ^{R} \rangle \langle
x_\lambda ^{R} |\right)\psi \rangle +{1\over 2}\left( |x_\lambda
^{L} \rangle \langle x_\lambda ^{L} |+ |x_\lambda ^{R} \rangle
\langle x_\lambda ^{R} |\right)\psi \rangle
\right],
\label{proection}
\end{eqnarray}
where matrices $\sigma^\pm_{ \lambda}$ are determined by
\begin{eqnarray}
\sigma^+_\lambda =\epsilon _{\lambda ij} \sigma_{ij};\
\sigma^-_{\lambda} =\epsilon _{\lambda ji} \sigma_{ij}, \ \lambda
, i,j =1,2,3; \label{4dPaulimatrix}
\end{eqnarray}
indexes $\kappa, \mu , \nu $
run over $s, \dot{s} $; $\sigma_{\mu \nu}= {\imath \over
2}[\gamma_\mu , \gamma_\nu ]$, $\gamma_\mu$
are the Dirac matrices \cite{Kaku}.
Hereinafter,  we will omit the sign $ \pm $. Since the skew-symmetric
tensor $\epsilon _ {\lambda jk} $ appears in the equation for $
\sigma _ {\lambda} $, the generalized Pauli matrixes  $ \sigma _
{\lambda} $ are   pseudo-vectors.
 Therefore, right- and left-helical  particles differ from one another.

It follows from the expansion (\ref {proection}) of the wave function  $ | \psi
\rangle $ in a series   that   the projection $P$   defined by the expression
\begin{eqnarray}
P|\psi\rangle = {1\over 2}\left( |x_\lambda ^{R} \rangle \langle
x_\lambda ^{R} |+ |x_\lambda ^{L} \rangle \langle x_\lambda ^{L}
|\right)\psi \rangle ,
 \label{proection1}
\end{eqnarray}
can be represented in the form
\begin{eqnarray}
P|\psi\rangle ={1\over 2}\left( |x_\lambda ^{R} \rangle \langle
x_\lambda ^{R} |+ |x_\lambda ^{L} \rangle \langle x_\lambda ^{L}
|\right)\psi \rangle
={1\over 2} \left\{ \sum_\mu |\xi_\mu \left> \right< \xi_\mu|\psi
\rangle + \sum_\lambda |\xi_\kappa \left> \left(\sigma_\lambda
\right)_{\kappa \nu} \right< \xi_\nu|\psi \rangle \right\}
\label{proection2}.
\end{eqnarray}

Let us prove  that the projector (\ref {proection2}) selects
states with a given orientation. The right side of the expression
(\ref{proection2}) in the representation $ \{| \xi_i \rangle \}_ i $ is written as
\begin{eqnarray}
P\langle \xi _\lambda |\psi \rangle ={1\over 2} \left\{ 1 +
\sum_\lambda \langle \xi _\lambda |\xi_\kappa \left>
\left(\sigma_\lambda \right)_{\kappa \nu}
\right.
\right\} \langle \xi _\nu |\psi \rangle \label{proection3}
\end{eqnarray}
 We can introduce a four-vector $s ^\lambda = \langle \xi _ \lambda
| \xi_\kappa \rangle $ describing a spin of the system. A
convolution of the vector $s ^\lambda $  with matrixes $
\sigma_\lambda $ is assumed on the right side in Eq.~(\ref
{proection3}). It follows from here, that $P=P (s) $ is determined
in $s $-representation by the expression
\begin{eqnarray}
P(s)={1\over 2}(1+ s^\mu \gamma_\mu )    \qquad  s^\mu =\{0,s^\lambda \}.
\label{proector(s)}
\end{eqnarray}
 The formula (\ref{proector(s)}) is a known expression for the
projection operator that selects  bispinors with a given orientation
in the rest frame \cite {Kaku}. Hence, it is possible to present
the projector $P $  in the form (\ref {proection2}).
The equality (\ref {proection2}) in a matrix form (\ref{presentation})
for the representation $ \Phi =X=X^R = {X^L} ^{\dagger} $ has the following form
\begin{eqnarray}
P|\psi\rangle ={1\over 2}\left( |x_\lambda ^{R} \rangle \langle
x_\lambda ^{R} |+ |x_\lambda ^{L} \rangle \langle x_\lambda ^{L}
|\right)\psi \rangle
={1\over 2} \left\{ \sum_\mu |\xi_\mu \left> \right< \xi_\mu|\psi
\rangle + \sum_\lambda |\xi_\kappa \left> \left(\sigma_\lambda
\right)_{\kappa \nu} \right< \xi_\nu|\psi \rangle \right\}
\label{proection4}
\end{eqnarray}
After left multiplication  on $ \langle x_\lambda ^R | $ the
  requirement of orthonormal basis $
\langle x_{\lambda_i} ^R | x_{\lambda_j} ^R \rangle =\delta_{ij} $ allows us to
rewrite the expression (\ref{proection4}) in the form
\begin{eqnarray}
  \langle x_\lambda ^{R}
|x_\lambda ^{L} \rangle \langle x_\lambda ^{L} |\psi \rangle
= \langle x_\lambda ^{R}|\xi_\kappa \left> \left(\sigma_\lambda
\right)_{\kappa \nu} \right< \xi_\nu |x_\lambda ^{L} \rangle
\langle x_\lambda ^{L}|\psi \rangle \label{proection5-0}
\end{eqnarray}
Let us introduce the designations
\begin{eqnarray}
x_\lambda = \langle x_\lambda ^{R}| x_\lambda ^{L} \rangle;\qquad
\xi_\kappa = \langle x_\lambda ^{R}|\xi_\kappa \rangle .
\label{reduction}
\end{eqnarray}
 Taking into account  the expression (\ref{reduction})
 after simple transformations the expression (\ref{proection5-0}) is written as
\begin{eqnarray}
   x_\lambda
 = \xi_\kappa \left(\sigma_\lambda \right)_{\kappa \nu}
\xi_\nu . \label{proection5}
\end{eqnarray}
From the above mentioned  it  follows  that to carry out  the
projection, coordinates should be changed  in the wave equation  (\ref {rel-eq3})
according to Eq.~(\ref {proection5})  and
the wave function $ \psi $  should be chosen so that its projection $P $
reads
\begin{eqnarray}
|\psi \rangle = P|\psi \rangle \label{degeration} .
\end{eqnarray}
Let us choose  a polar coordinates system $\left.\{\rho_i,
\chi_i\}\right|_{i=1}^2$ which obeys  $\rho
_i=\sqrt{\xi_i^2+\xi_{i+1}^2}, \ i=1,2$, and $\chi_i $ is a polar angle in the plane
$ \{\rho_i, \chi_i\}, i=1,2$. Then
the condition (\ref{degeration})  takes the form
\begin{eqnarray}
\langle \rho_1,  \rho_2, \chi_1,\chi_2|\psi \rangle = \langle
\rho_1,  \rho_2, \chi_1, \chi_2| P|\psi \rangle = \langle \rho_1,
\rho_2,  \chi_2= f(\chi_1)|\psi  \rangle \label{degeration1} ,
\end{eqnarray}
where $f(\chi_1)$ is a function of
the polar angle $\chi_1$. %
The above-stated allows us to write  the equation (\ref {rel-eq3})
in the three-dimensional physical space.

Next, we examine  a case of the particle with an integer-valued spin in
the rest reference frame.

\section{
A motion equation  for a particle with an integer-valued spin}

We know that in  the rest frame the spin of a system can be described by a
three-vector  $ s ^\mu $ as
\begin{eqnarray}
s^\mu =\{0, s_1,  s_2,  s_3\}. \label{rest-spin}
\end{eqnarray}
Then, the vector $ \vec \xi _ {s} $ in the equation (\ref{rel-eq3}) is determined
on ordinary $c $-numerical space, i.e.
space with commutation  relations. Since components of the vector
$ \vec \xi _ {s} $ is ordinary $c $-number, the commutators in the
last summand vanish. Hence, we get
\begin{eqnarray}
  \left(
-{\partial ^2\over \partial  \xi 
_{\dot{s}}  \partial \xi _s }
-Z^2 + \omega ^2 \left( \overline{\xi}_{\dot{s}}
\xi_{s} 
\right) \left|\Phi \rangle \langle \Phi \right|
 \right)\psi\rangle  = 0.
  \label{oscillator}
\end{eqnarray}
Here $ \overline {\xi} _ {\dot {s}} $ is a transposed spinor
conjugated to the spinor $ \xi _ {s} $. The obtained equation
(\ref {oscillator}) is the equation for an isotropic harmonic
quantum oscillator with the mass $m = {1\over 2} $ in
two-dimensional space with complex coordinates $ \xi_s \stackrel
{def} {= \hspace {-0.8mm} =} \langle \Phi=X | \xi_s \rangle \equiv
\vec \xi_s $, $s=1,2 $ that are  components of bispinor in the
representation $ \Phi =X $, determined on ordinary $c $-numerical
space.

In the next section we will project a trajectory of a  particle on
three-dimensional space.

\section{
Transformation to space-time coordinates}

 Taking into account the condition (\ref {rest-spin}) we calculate
the square  of module of the expression (\ref {proection5}):
\begin{eqnarray}
x^\lambda x_\lambda =  \left| x \right|^2
 =
  \left| \left(\
\begin{array}{c}
 \xi_s\\
\xi_{\dot{s
}}
\end{array}
 \right)^{\dagger}
 \left(
\begin{array}{cc}
0 &  \sigma_i\\
-\sigma_i & 0
\end{array}
 \right)
\left( \begin{array}{c}
 \xi_s\\
\xi_{\dot{s
}}
\end{array}
 \right)
\right|^2
\nonumber \\[0.2cm]
= |\xi |^2\left(\
\begin{array}{c}
 \xi_s\\
\xi_{\dot{s
}}
\end{array}
 \right)^{\dagger}
 \left(
\begin{array}{cc}
0 &  \sigma_i^2\\
\sigma_i^2 & 0
\end{array}
 \right)
\left(  \begin{array}{c}
 \xi_s\\
\xi_{\dot{s
}}
\end{array}
 \right)=|\xi |^4
. \label{square-proection5}
\end{eqnarray}
Here $ | \xi | ^2 =\sum _ {\nu =1} ^4 \xi^2_\nu $,
the Dirac's representation of matrices $ \gamma ^ \mu, \ \mu
=0,1,2,3 $ was
chosen; and it was taken into account  that $ \sigma_i^2=1 $, where
$ \sigma_i $ are the Pauli matrices. It follows from the obtained
expression (\ref {square-proection5})  that
\begin{eqnarray}
\xi^2_1+\xi^2_2+\xi^2_3+\xi^2_4=r \label{square-proection6}
\end{eqnarray}
where $r = | x | $. According to the expression
(\ref{square-proection6}) the following replacement  of
variables $ \xi_\kappa \to x_\lambda $ in the motion equation
(\ref{oscillator})  may be performed:
\begin{eqnarray}
\xi^2_1+\xi^2_2= {r\over 2}+x_3=u; \qquad \xi^2_3+\xi^2_4={r\over
2}-x_3=v \label{change}
\end{eqnarray}
As seen  from  Eq.~(\ref {change})   it is necessary to
pass  to the polar coordinates $\left.(\rho_i,\chi_i)\right|_{i=1}^2$
\begin{eqnarray}
\rho_i = \sqrt{\xi_i^2+\xi_{i+1}^2},\qquad  \chi_i=\arctan \left(
{\xi_{i+1}\over \xi_i } \right),\ i=1,2 . \label{change1}
\end{eqnarray}
In these coordinates the square of an arc element determined  by
\begin{eqnarray}
ds^2_1=d\rho_1^2+\rho_1^2 d\chi_1^2={1\over u}du^2+u d\chi_1^2;\\
ds^2_2=d\rho_2^2+\rho_2^2 d\chi_2^2= {1\over v}dv^2+vd\chi_2^2
\label{arc}
\end{eqnarray}
where variables $u, v $ are defined  by the expression (\ref{change}).
Evidently, that in the coordinates $ \xi_s, \xi _ {\dot s} $
the equation (\ref{oscillator}) is solvable by a separation of
variables. Therefore, Laplacian can be written in the form $
\Delta = \Delta_s + \Delta _ {\dot s } = \sum _ {i=1} ^2 \Delta_i $.
To get the motion equation    in the polar coordinates $ \rho_i,
\chi_i, \ i=1,2 $ we  use a theorem according to which the Laplacian
$ \Delta_i, \ i=1,2 $ in
curvilinear orthogonal coordinates $q _ {1i}, q _ {2i} $ has the form
\begin{eqnarray}
\Delta_i   = {1\over h_{1i}h_{2i}}\left\{ {\partial \over \partial
q_{1i}}\left({h_{2i}\over h_{1i}} {\partial  \over \partial
q_{1i}}\right) + {\partial \over \partial
q_{2i}}\left({h_{1i}\over h_{2i}} {\partial  \over \partial
q_{2i}}\right) \right\}, \label{Lame-coeffic}
\end{eqnarray}
where $h _ {1i}, \ h _ {2i} $ are the Lame coefficients in the expression
for the square of the differential of  arc-length
\begin{eqnarray}
ds^2_i=h_{1i}^2 dq_{1i}^2+h_{2i}^2 dq_{2i}^2 \label{general-arc}.
\end{eqnarray}
By applying this theorem to our case, after the simple transformations
we gain the Laplacian in coordinates $(u,v,\chi_1,\chi_2)$:
\begin{eqnarray}
\Delta _{u,v,\chi_1,\chi_2} =  \left\{ {\partial \over \partial
u}\left(u {\partial  \over \partial u}\right) + {\partial \over
\partial v}\left(v {\partial \over \partial v}\right) +{1\over u}
{\partial^2   \over \partial\chi_1^2}+ {1\over v}{\partial^2
\over \partial\chi_2^2} \right\} \label{Laplacian}.
\end{eqnarray}
Since   the degenerate discrete representations of the group \textsf{SO(4,2)}
coincide with the projectors  defined by subalgebra \textsf
{su (1,1)} \cite{barut}, one has a relation $m_1=m_2=m $ in the
expression for a wave function $ \psi = U (u) e^ {\imath
m_1\chi_1} V (v) e ^ {\imath m_2\chi_2} $. Therefore the wave
function is represented as
\begin{eqnarray}
\psi = U(u)V(v)e^{\imath m \chi }. \label{Laplacian1}
\end{eqnarray}
Here
\begin{eqnarray}
 \chi =\chi_2 +\chi_1 .
\end{eqnarray}
Hence, as a manifold $ \{u, v, \chi, \eta \} \in \mathbb{R}^4
$, $ \eta = \chi_2 -\chi_1 $, on which   motion of the particle is
examined, one can choose  such a submanifolds being a section
\begin{eqnarray}
\eta = 2\pi a\ n, \qquad n=1,2,\ldots \label{manifold}
\end{eqnarray}
where $a$ is a constant.

Let us introduce cylindrical coordinates in three-dimensional physical
space $z,\rho , \phi$:
\begin{eqnarray}
z =  x_3 = {u-v\over 2} , \quad \rho = \sqrt{uv} \quad \phi ={\chi \over 2}
 \label{Laplacian1-1} .
\end{eqnarray}
The polar angle $ \phi $ varies within the limits from $ 0 $ to
$2\pi $, because the   angle $ \chi $ varies   from $ 0 $ to $4\pi $.
 Eqs.~(\ref{manifold}), (\ref{Laplacian1-1}) allows to make the following
replacement
\begin{eqnarray}
{\partial \over \partial \chi_2} \left( {\partial \over \partial
\chi_1} \right) \Rightarrow {1\over 2}{\partial \over \partial
\phi } \label{derivative-change} .
\end{eqnarray}
Substitution of Eqs.~(\ref{Laplacian}, \ref{derivative-change})
into Eq.~(\ref{oscillator}) gives
\begin{eqnarray}
  -\left\{
{\partial \over \partial u}\left(u {\partial\psi \over \partial
u}\right) + {\partial \over \partial v}\left(v {\partial\psi \over
\partial v}\right) +{1\over 4} \left( {1\over u} + {1\over
v}\right){\partial^2 \psi \over \partial\phi^2} \right\}
-\left( Z^2 - \omega ^2 (u+v) 
\right)
 \psi   = 0.
  \label{3d-oscillator}
\end{eqnarray}
Dividing the equation (\ref {3d-oscillator}) on $u+v $ and
 multiplying on 4,  one gets
\begin{eqnarray}
  -{4\over u+v}\left\{
{\partial \over \partial u}\left(u {\partial\psi \over \partial
u}\right) + {\partial \over \partial v}\left(v {\partial\psi \over
\partial v}\right) +{u+v \over 4uv}
 {\partial^2 \psi \over \partial\phi^2}
\right\}
-4\left( {Z^2 \over u+v} - \omega ^2  
\right)
 \psi   = 0.
  \label{3d-oscillator1}
\end{eqnarray}
Denote
\begin{eqnarray}
  \xi^2=\xi^2_1+\xi^2_2;\ \eta^2=\xi^2_3+\xi^2_4
\label{Laplacian3},
\end{eqnarray}
Because
\begin{eqnarray}
dz^2+d\rho^2  =(\xi^2+\eta^2)(d\xi^2+d\eta^2)= {1\over 4}(u+v )
\left( {1\over u}du^2+{1\over v}dv^2 \right)
 \label{Laplacian3-1},
\end{eqnarray}
the square  of arc element
\begin{eqnarray}
ds^2=dz^2+d\rho^2  +\rho^2 d\phi^2
 \label{arc3-1}
\end{eqnarray}
in cylindrical 3d-frame is equal  to
\begin{eqnarray}
ds^2=  {u+v\over 4u}   du^2+{u+v\over 4v}dv^2+uv d\phi^2
\label{arc3-2}
\end{eqnarray}
in (u,v)-coordinates. Thus, the Laplace operator in 3d-space has the form
\begin{eqnarray}
\Delta  _{q_1,q_2,q_3} = {1\over h_{1 }h_{2 }h_3}\left\{ {\partial
\over \partial q_{1 }}\left({h_{2 }h_3\over h_{1 }} {\partial
\over \partial q_{1 }}\right) + {\partial \over \partial q_{2
}}\left({h_3 h_{1 }\over h_{2 }} {\partial \over \partial q_{2
}}\right) + {\partial \over \partial q_{3 }}\left({ h_{1 }h_{2
}\over h_3} {\partial  \over \partial q_{3 }}\right) \right\},
\label{3d-Lame-coeffic}
\end{eqnarray}
where coefficients $h_{1 }$, $h_{2 }$, $h_3$ determine
the square of the arc element as
\begin{eqnarray}
ds^2 =h_{1 }^2 dq_{1 }^2+h_{2 }^2 dq_{2 }^2+h_{3 }^2 dq_{3 }^2
\label{3d-general-arc}.
\end{eqnarray}
From the expression for the Laplace operator
\begin{eqnarray}
 \Delta _{u,v,\phi} =  {4\over u+v}\left\{
{\partial \over \partial u}\left(u {\partial  \over \partial
u}\right) + {\partial \over \partial v}\left(v {\partial  \over
\partial v}\right) +{u+v \over 4uv}
 {\partial^2   \over \partial\phi^2}
\right\}
  \label{3d-Laplas1}
\end{eqnarray}
appearing in Eq.~(\ref{3d-oscillator1}) it follows that
\begin{eqnarray}
h_1={1\over 2}\sqrt{{u+v\over u}},\quad h_2={1\over
2}\sqrt{{u+v\over v}}, \quad h_3= \sqrt{u v }.
\label{3d-Lame-coeffic1}
\end{eqnarray}
Comparing expressions (\ref {3d-Lame-coeffic1}) and (\ref
{arc3-2}) it may be concluded  that the Laplacian (\ref {3d-Laplas1})
was written in parabolic 3d-frame $u, v $:
\begin{eqnarray}
u=R+z, \qquad v=R-z, \label{parabolic-system}
\end{eqnarray}
in which the  square of the arc element has the form (\ref{arc3-2}).
Comparing the expression (\ref{change}) with the expression
(\ref{parabolic-system}) one obtains that
\begin{eqnarray}
 R ={r\over 2},
\label{sphere-system}
\end{eqnarray}

\section{Derivation of wave equations for charged vector-boson on non-compact group
}

According to expressions (\ref {arc3-1}, \ref {3d-Lame-coeffic},
\ref {3d-general-arc}, \ref{3d-Lame-coeffic1}, \ref{sphere-system}) it follows
that the motion equation   (\ref{3d-oscillator1}) in the spherical frame can be
rewritten as
\begin{eqnarray}
- \Delta \psi -  {2Z^2 \over R} \psi =-4\omega^2
 \psi   .
  \label{3d-oscillator2}
\end{eqnarray}
After elementary  substitution $4\omega^2 \Rightarrow
E_1^2-m_1^2$, $Z^2 \Rightarrow \alpha (E_1+m_1) $ one gets
\begin{eqnarray}
\left[ -\nabla^2 - \left(E_1 +{  \alpha \over R}\right)^2+
\left(m_1 -{\alpha \over R}\right)^2 \right]
 \psi  = 0,
  \label{oscillator2}
\end{eqnarray}
which, in principle is the Klein-Gordon equation, as it must,
with a scalar potential $ \varphi (R) = -\imath \alpha /R
$ and a vector potential $A (R) = -\imath \alpha /R $, where $A
(R) $ is  a displacement of the $R $-th component of the gradient $
\vec \nabla $. This results  from the form of the Laplacian in the spherical
coordinates and the facts that
$ \psi \equiv \{\psi_i\}_{i=1}^3\stackrel{def}{=\hspace{-1mm}=}\vec \psi $ and
the spherical symmetry of a problem yields
$ \vec A\cdot \vec
\nabla _R\psi \equiv \vec A\cdot (\vec \nabla _R\times\vec \psi) =0 $;  where
$\vec \nabla _R$ is a radial component of gradient $\vec \nabla$.
Therefore, as a result, the equation (\ref{oscillator2}) can be rewritten as
\begin{eqnarray}
\left[ \hat P^2  - \left(E_1 +{  \alpha \over R}\right)^2+
m_1 ^2 \right]
 \psi ^* = 0, \qquad \hat P=\tau_3\left(P_R+
\imath {{\cal M}\over R},
\right),  \qquad {\cal M}=-{\tau_1\over \sin\theta } p_\phi +\tau_2 p_\theta
  \label{oscillator2complect}
\end{eqnarray}
where $ \psi ^*$ is the wave function of a particle with an integer
spin, which is determined through the generalized spherical function
with an integer spin and results from the  transformation
of the original wave function similar  to fermions \cite{Fok};
the wave function $ \psi ^*$ is an eigenfunction of the operator $
{\cal M} $: $ {\cal M} \psi = m_1\psi $; $ \psi  = \psi ^*/R$;
matrices $ \tau_i $,
$i=1,2,3 $ are generators of the   rotation group \textsf{SO(3)}
in a matrix form, which  correspond  to the generalized Pauli matrices
for the spherically symmetric Dirac equation; $ \hat P $ is
 the relativistic momentum operator obtained by adding  components of
the vector-potential   $ \vec A = \{A (R), 0,0 \} $ to the
ordinary momentum operator which is   a derivative (tangential
vector):
 $P_R=p_R+A(R)$,
$p_R=-\imath\hbar {\partial \over \partial R} $, $p_\theta
=-\imath\hbar {\partial \over \partial \theta } $, $p_\phi
=-\imath \hbar {\partial \over \partial \phi } $.

Moreover, it can be seen  that when taking into account  the explicit
 expression for  eigenvalues $Z^2 $ of the Hamiltonian of two-dimensional
harmonic oscillator (\ref{oscillator}) with the mass $m = {1\over 2} $ and
eigenfrequency $\omega _0=2\omega $:
\begin{eqnarray}
Z^2= 2\omega (n_r+l+1),\ \hbar =1
  \label{oscillator3}
\end{eqnarray}
then Eq.~(\ref{3d-oscillator2}) can be rewritten as
\begin{eqnarray}
- \Delta \psi -  {2Z^2 \over R} \psi = E_n
 \psi   , 
  \label{3d-oscillator3}
\end{eqnarray}
where
\begin{eqnarray}
 E_n =-{Z^4\over (n_r+l+1)^2}, \ m={1 \over 2 },\ \hbar =1 ,
  \label{3d-oscillator3-1}
\end{eqnarray}
$n_r , l $ are quantum numbers of the harmonic oscillator with the
Laplacian (\ref {Lame-coeffic}). Let  us assume that
$2Z^2\Rightarrow e^2 $. If $e $  is supposed to be  the charge of
an electron then the equation (\ref {3d-oscillator3}) is nothing
but the equation on eigenfunctions with eigenvalues $E_n $   of a
hydrogen-like atom.

\section{Asymptotic solution}
Apparently, one can write for  the relativistic boson (\ref {oscillator2}) that
$(E_1^2-m_1^2)n^2=\alpha^2(E_1+m_1)^2$ and ${Z^2\over n^2}={E_1-m_1\over \alpha}$.
 Therefore, the relativistic boson   has the following
spectrum $E_1=m_1{\left(1+{\alpha^2\over {n^*} ^2}\right)
\over\left(1-{\alpha^2\over {n^*}^2}\right)}$ with a mass spectrum
$m_1={\tilde Z^2\over 2}\left(1-{\alpha^2\over {n}^2}\right)$, 
where $n =n_r+l+1$, $n^*=n^*_r+l^*+1$, $\tilde Z^2={Z^2\over \alpha}$.
Thus, Eq.~(\ref{oscillator2}) describes a compound physical system.
Let us choose $ \alpha = \imath \gamma, \ \gamma \in \mathbb {R}
$, that yields the Coulomb potential. Then, the equation (\ref{oscillator2}) allows
relativistic corrections to the
spectrum of the hydrogen-like atom   in the non-relativistic limit $n
(n ^ *)\to \infty $  to be found. With the above mentioned  consideration, the
spectrum in this case has the form
\begin{eqnarray}
 E_1\approx {\tilde Z^2\over 2}\left(1-{\gamma ^2\over {n^*}^2}\right)
 { \left(1-{\gamma ^4 \over 4 n ^4}\right)}
\end{eqnarray}
if the spectrum is assumed to  consist of two  spectral series:
odd one $\{ n^*\}$ and even one
$ \{ n \}\to  \{ 2n \} $ and the  series are close:
$\gamma ^2/{n^*}^2 - \gamma ^2 / 4 n ^2 =\overline \epsilon $, $\overline
\epsilon \ll 1$.
Since the  series are close,
\begin{eqnarray}
 {n^*}^2\approx
 n ^2+\epsilon , \qquad  |\epsilon |/n ^2 \ll 1  .
\end{eqnarray}
Let us choose $\epsilon $ as
\begin{eqnarray}
  \epsilon = 2(n - |k|)(\sqrt{k^2-\gamma^2} - |k|)\approx
  -(n - |k|){\gamma ^2\over |k|},
\end{eqnarray}
where $k= -l,l+1$. %

Introduce designations: $\tilde Z^2 \Rightarrow m $, where $m $ is
a particle mass. Then, expanding  into a series on $ {\gamma ^2 (n
- |k |)\over |k |} $, we gain accurate to terms of  the order $m\gamma ^8$
\begin{eqnarray}
 E_1\approx {m\over 2} -{m\gamma ^2 \over 2{n }^2} - {m \gamma ^4\over 8 n ^3}
 { \left({4\over   |k|  }-{3\over   n  }\right)}-
 {m \gamma ^6\over 8 n ^4}
 { \left({3\over   n^2  }-{8\over  n  |k|  }+{4\over  k^2  }\right)}
 +O(\gamma^8)
 \label{boson-energy}.
\end{eqnarray}
From the formula (\ref{boson-energy}) it follows that the first
term yields the relativistic rest energy of the compound system
considerd with a reduced mass $m/2 $. The second term yields
the Rydberg formula. The third term is a relativistic correction
which describes the fine structure of the hydrogen-like atom lines with
the fine structure constant $ | \gamma | $ in the system of units:
$e=1 $, $ \hbar=1 $, $c=1 $. The fourth term is a relativistic
correction which   yields the computational method used.

\section{Symmetry of wave equations for a particle with integer-valued spin
 }
Now, let us find  symmetry groups of  the wave equations
(\ref{oscillator2}) and (\ref{3d-oscillator3}).
With this goal one defines on bispinor $(\xi_s, \xi_{\dot s} )^T$
a creation operator $\hat a^+$ and a annihilation operator $\hat a$ as
\begin{eqnarray}
\label{annihilation}
\hat a =
\left(
\begin{array}{c}
a_s\\
a_{\dot s}
\end{array}
\right)=
\left(
\begin{array}{c}
\sqrt{\omega}\left(
\xi_s+{1\over \omega}{ \partial \over \partial \xi_{\dot s}}
\right)\\[0.2cm]
\sqrt{\omega}\left(
\xi_{\dot s}+{1\over \omega}{ \partial \over \partial \xi_s}
\right)
\end{array}
\right),
 \\[0.4cm]
 \label{creation}
 \hat a^+ =
\left(
\begin{array}{c}
a^+_s\\
a^+_{\dot s}
\end{array}
\right)^T=
\left(
\begin{array}{c}
\sqrt{\omega}\left(
\xi_{\dot s}-{1\over \omega}{ \partial \over \partial \xi_{ s}}
\right)\\[0.2cm]
\sqrt{\omega}\left(
\xi_{ s}-{1\over \omega}{ \partial \over \partial \xi_{\dot s}}
\right)
\end{array}
\right)^T
 \end{eqnarray}
where the sign $T$ denotes the transposition operator.
Then, we find accurate up to a sign operators commuting with the
hamiltonian $H$ from Eq.~(\ref{oscillator}):
\begin{equation}
\begin{split}
&M_\lambda   = (\sigma_\lambda)_{{\dot s}t}a_t a_{\dot s}, \qquad
 M^+_\lambda = (\sigma_\lambda)_{{\dot s}t}a_s^+ a_{\dot t}^+, \\
&N^a_\lambda = (\sigma_\lambda)_{s t}a_s^+ a_{t}, \qquad
N^b_\lambda = (\sigma_\lambda)_{{\dot s}\dot t}a_{\dot t}^+ a_{\dot s}, \\
&P           = a_s a_{\dot s}, \qquad\qquad  \ \ \  P^+=a^+_s a^+_{\dot s},\\
&H           = 2+a^+_s a_s+a^+_{\dot s} a_{\dot s}.
\end{split}
\label{algebra-basis}
\end{equation}
Using commutation relations for the annihilation (\ref{annihilation}) and
the creation (\ref{creation}) operators  it can be shown that fifteen
operators (\ref{algebra-basis})
will constitute a basis set of generators for the algebra of a
symmetry group similar to the Lie algebra  of the group \textsf{SO(4,2)} or
\textsf {SO(6)}.   The Lie group \textsf{SO(4,2)} is
non-compact while the  Lie group \textsf{SO(6)} is the compact group.
According to "no-go"\ -theorem \cite{Kaku}, \cite{Barut-V.2-p.296}
 taking  place for boson degrees of
freedom, there are no finite-dimensional  unitary
representation of the noncompact Lie group. This suggests  that
any nontrivial union of the Poincar\'e and an interior
symmetry group yields an $S$ matrix which is equal to $1$. The equation
(\ref{oscillator2}) or (\ref{oscillator2complect}) describes a particle
 which wave function    is transformed on the infinite-dimensional representation
of the noncompact Lie group. From here and from the "no-go" \   theorem
it follows that the obtained relativistic equation (\ref{oscillator2complect}) has
the symmetry of \textsf {SO(4,2)} and is
reasonable for the description in an arbitrary reference frame.

\section{Conclusions}
The equation (\ref {3d-oscillator3}) is the Schr\"odinger equation
for the hydrogen-like atom. The equation (\ref {3d-oscillator3})
also should describe locally a physical system having a symmetry
similar to the noncompact symmetry group \textsf {SO(4,2)}.
 Fock \cite{Fok1} was first to indicate invariance of the Schr\"odinger equation
for the
hydrogen-like atom with respect to four-dimensional rotations.
But the equation (\ref{3d-oscillator3}) does not contradict to
"no-go"\ -theorem only in the case if it describes a particle,
which wave function is transformed according to representation of
a compact group similar to the group \textsf{SO(4,2)}.
The group \textsf {SO (4,2)} is similar to the group \textsf{SO(6)}.
The generators of these groups differ by signs of metric
multipliers \cite{Barut-V.1p.20-V.2p.44;Kaku-p.53}. Hence
we conclude, that the equation (\ref{3d-oscillator3}) describes a
particle, which wave function  is transformed on representation of
the compact group \textsf {SO (6)}.
This description is acceptable  to the rest reference frame  when it
is possible to neglect the Lorentz boost transformations of the group
\textsf {SO (4,2)}.

Thus, within the framework of the projection
operator method two wave equations for a vector charged boson are
obtained, one of which is precise (equation (\ref {oscillator2})),
and the second equation (the equation (\ref {3d-oscillator3}))
describes a motion of a vector charged boson approximately. The
hydrogen-like atom is the compound quantum system with an integer
spin and in the stationary case that does not take into account the
Lorentz boost transformations, the replacement of a dynamic symmetry group
\textsf{SO(4,2)} on \textsf{SO(6)} yields the non-relativistic
Schr\"odinger equation for the hydrogen-like atom. On this basis, we get the
conclusion  that the obtained relativistic corrections (\ref{boson-energy})
describe the fine structure of the hydrogen-like atom
lines with precision to about $   \gamma   ^ 8 $.
In general, infinite-component wave equations have also been examined by Nambu
\cite{Nambu}   to describe the hydrogen-like  atom.

\end{document}